\documentclass{article}

\usepackage{amssymb}
\usepackage{amsmath}
\usepackage{graphicx}
\usepackage{color}
\usepackage{url}

\newcommand{\hi}[1]{{#1}}

\begin{document}

\author{
Mario Frank$^{\sharp}$, Ralf Biedert$^{\dagger}$, Eugene Ma$^{\sharp}$, Ivan Martinovic$^{\flat}$, Dawn Song$^{\sharp}$
\\
{
\footnotesize
$^{\sharp}$UC Berkeley, 
$^{\dagger}$German Research Center for Artificial Intelligence (DFKI) GmbH,    
$^{\flat}$University of Oxford
}
}

\title{Touchalytics: On the Applicability of Touchscreen Input as a Behavioral Biometric for Continuous Authentication}

\maketitle

\begin{abstract}
We investigate whether a classifier can continuously authenticate users based on the way they interact with the touchscreen of a smart phone.
We propose a set of 30 behavioral touch features that can be extracted from raw touchscreen logs and demonstrate that different users populate distinct subspaces of this feature space.   
In a systematic experiment designed to test how this behavioral pattern exhibits consistency over time, we collected touch data from users interacting with a smart phone using basic navigation maneuvers, i.e., up-down and left-right scrolling.  
We propose a classification framework that learns  the touch behavior of a user during an enrollment phase and is able to accept or reject the current user by monitoring 
 interaction with the touch screen.
The classifier achieves a median equal error rate of 0\% for intra-session authentication, 2\%-3\% for inter-session authentication and below 4\% when the authentication test was carried out one week after the  enrollment phase. While our experimental findings disqualify this  method as a standalone authentication mechanism for long-term authentication, it could be implemented as a means to extend screen-lock time or as a part of a multi-modal biometric authentication system.
\end{abstract}

\section{Introduction} Most methods for authenticating users on desktop
computers or mobile devices define an entry point into the system.  Typically,
the user faces a password challenge and is granted access only if she inputs
the correct password. While such entry-point based methods dominate the
authentication schemes today, they have flaws from both usability and security perspectives. 
From a usability perspective, traditional authentication schemes are inconvenient
because users must focus on the authentication step every time they begin
interacting with their device.  Such inconvenience is amplified under the usage
pattern of mobile devices, since they are more frequently accessed, and
each use is typically shorter.  Authentication with a PIN or
secret gesture is too cumbersome for short bursts of activity, such as briefly
checking one's email or reading an SMS. Hence, users often choose simple and weak secrets, increase the screen lock timeouts of their devices, or completely disable unlock~\hi{~\cite{symantec,sophos}}. 
Recent studies have demonstrated how simple attacks such as smudge attacks
\cite{smudge} can break entry-point authentication schemes. Furthermore, the device cannot detect intruders
after the authentication step is performed successfully.

\begin{figure}[t]
   \centering
   \includegraphics[width=0.8\columnwidth]{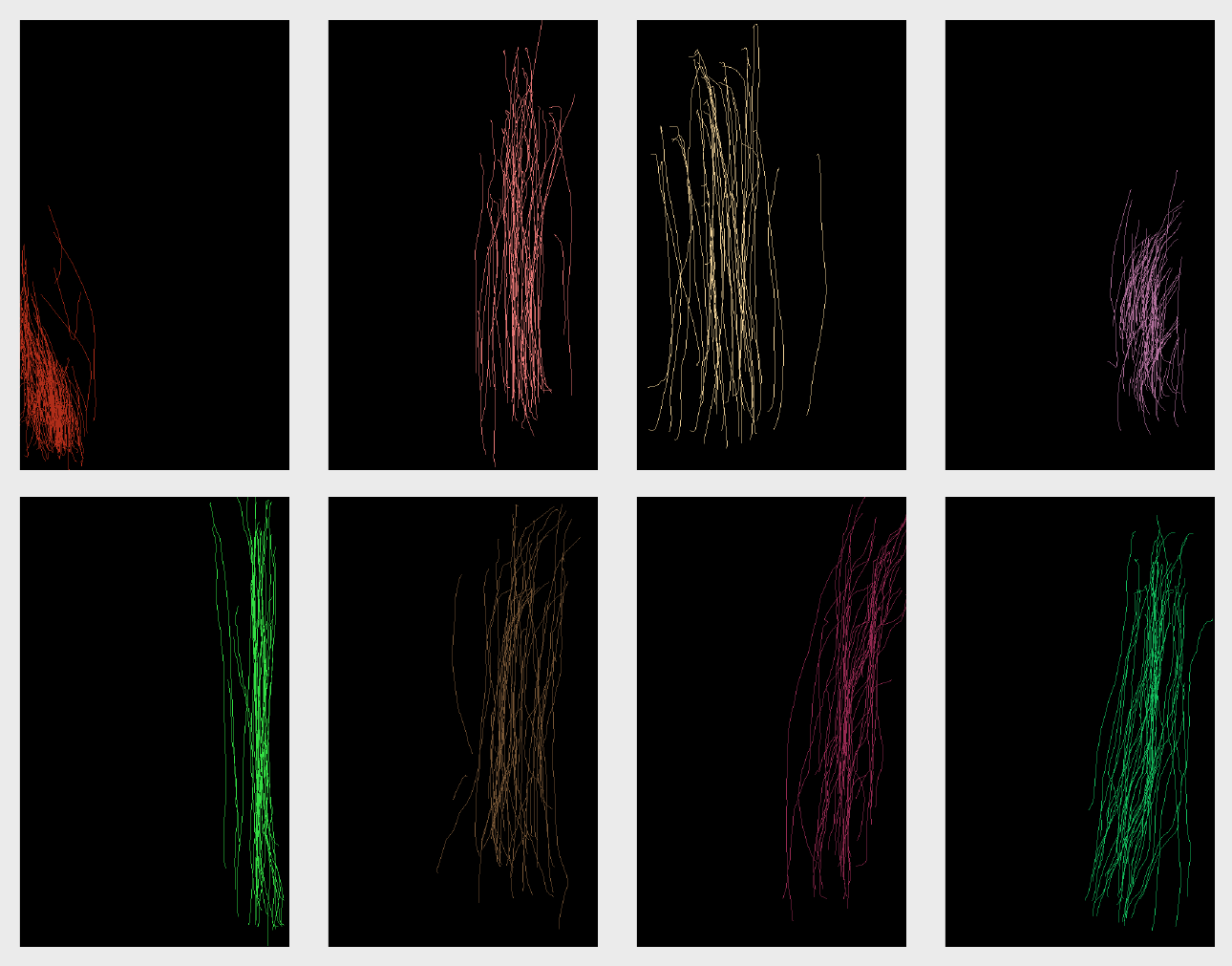} 
   \caption{\hi{Each user's} interaction behavior on touchscreens can be quite unique. This figure depicts strokes recorded from eight different users, each reading three different texts on an Android phone. Geometric patterns that discriminate the users from each other are already apparent. Other differences might come from different stroke timing, pressure, and area covered on screen. In this paper, we investigate to what extent such touch features can be used for user disambiguation and authentication.}
   \label{fig:example}
\end{figure}

Mobile devices are at a higher risk of loss or theft compared to desktop computers\hi{~\cite{losstheft}}. Continuous or implicit authentication approaches
would provide an additional line of defense, designed as a non-intrusive and passive security countermeasure. 
 Such approaches monitor the user's interaction with the device, and ideally, at every point in time (or at least with a high frequency) the system estimates if the legitimate user is using the device. Hence, a continuous authentication method can either complement entry-point based authentication methods by monitoring the user after a successful login or, if the method satisfies particular accuracy requirements, it could even substitute entry-point based authentication.

Although there is a growing body of literature about keystroke dynamics or
mouse dynamics for continuous authentication, there is surprisingly little work
on continuous authentication for touchscreen devices. The growing popularity
of mobile devices -- in 2011, more smart phones were sold than desktop PCs and
notebooks combined \cite{Canalysreport} -- increases the value of research on their security mechanisms.
Specifically, to the best of our knowledge, there is no existing method for continuous
authentication based on touch biometrics (i.e., without requiring a dedicated
activity of the user). One reason might be the difficulty of extracting a set of
sufficiently discriminative features from touch data, because atomic
navigation behavior mostly consists of simple and short movements (see
Figure~\ref{fig:example}).

In this paper, we lay foundational work  for continuous authentication schemes
that rely on touchscreen input as a data source. We investigate if it is
possible to authenticate  users while they perform basic navigation steps on a touchscreen device and without any dedicated and explicit security action that requires
attention from the user.  Our goal is to analyze how robustly such schemes
operate and if they are sufficiently reliable to be used on commodity devices.

Our contribution is a classification framework that serves as a
proof-of-concept for touch-based behavioral biometric authentication.  We
propose a set of 30 behavioral features that can be extracted from the touch
screen input of commodity mobile devices.  \hi{We designed} experiments that
let users interact with touchscreens in different sessions and with different
tasks, and demonstrate that our features are highly discriminative. Along
the way, we discuss design decisions and usage scenarios for such a continuous
authentication method. Our study provides insights in the operational
modes and scenarios that are permitted given the accuracy of the proposed
method. All data collected for this paper is available online\footnote{ \url{http://www.mariofrank.net/touchalytics/index.html} }.

\section{\hi{Related work}}

Biometrics, i.e., using human characteristics for identification and verification purposes has been an active research area for many years \cite{1634356}. Conventionally, it is divided into two categories \cite{Yampolskiy:2008:BBS:1389509.1389515}: physiological and behavioral biometrics. While physiological biometrics rely on static physical attributes, such as fingerprints, hand geometry, facial features, or DNA, behavioral biometrics aim at identifying invariant features of the human behavior during different activities such as speaking,  typing, or walking \cite{5638036}.

Early behavioral biometrics have been based on keystroke dynamics and mouse movements. In \cite{spillane75}, the authors introduce a keyboard system that captures timing and pressure characteristics to identify users  based on entering telephone numbers and PIN inputs. 
The work in \cite{Obaidat} improved the original system along different dimensions. The average error rates vary between 5\% - 15\%  depending on the input data \cite{Clarke2007109} (in this case, the length of telephone no. and PIN).
Keystroke dynamics gained lots of popularity through \cite{Monrose}, where it was used to augment password authentication with additional security. 
A survey on the large body of literature on authentication with keystroke dynamics is given in \cite{Joyce}.

In \cite{Koreman}, the authors use multi-modal biometrics composed of voice, face, and signature data for authentication on mobile phones. The goal is to enable legally binding contracts to be signed. While the face verification shows very high EER, around 28\%, the EER of voice and signature are around 5\% and 8\%, respectively. The fusion of the three biometrics decreases the EER to 	2\%, yet the price to be paid is the highly intrusive authentication procedure where the user needs to sign, read, and enter a PIN-based password.

Many authentication schemes require such an explicit user interaction, like typing a particular pass-phrase, or entering in a numerical PIN or secret gesture \cite{Oorschot}. 
However, there is a growing  body of literature that aims at continuous or implicit authentication. For instance, in \cite{niinuma:76670L}, the authors describe a continuous user authentication scheme based on monitoring user's face and color of clothing using a web cam. In \cite{touchPad}, keystroke dynamics were used to authenticate 10 users entering digits on on a 3.6$\times$7$cm^2$ touch pad.
Particularly suitable for continuous authentication are mouse dynamics as a behavioral biometric.
While in some studies on mouse biometrics the problem was identified too unreliable for authentication \cite{Pusara:2004:URV:1029208.1029210} others report high accuracies \cite{Nakkab,Gamboa,ZhengCCS2011}. For instance, in \cite{ZhengCCS2011} the EER is as low as 1.3\% when taking mouse actions between 20 successive clicks into account. The best accuracy has been reported in \cite{Nakkab} with a FAR of 0.36\% and a FRR of 0\%, although it has been suspected  that this result was influenced by recording the data on a different computer for each user \cite{Jorgensen:2011:MDB:1966913.1966983}.
In \cite{Ahmed}, mouse and keystroke dynamics have been combined to a multi-modal authentication system.  A critical discussion of mouse-based approaches together with a list of experimental pitfalls of continuous authentication is provided in \cite{Jorgensen:2011:MDB:1966913.1966983}. This work inspired some of our experimental design decisions.
Generally, mouse data and touch pad data is different from touch screen data in that a touch screen has no visible pointer. With a mouse, one continuously moves a pointer and uses clicks to carry out actions. Moving the finger to another position is invisible to a touch screen. This reduces the rate at which data is available to the system. On the flip side, all interactions with a touch screen correspond to an intended action, possibly making the gestures less random than for a mouse.

\hi{Touchalytics resembles the field of on-line signature authentication or on-line signature verification~\cite{Lee:1996:ROH:232678.232685,Jain20022963} in that it extracts temporal features of human gestures on planar surfaces. In~\cite{Lee:1996:ROH:232678.232685}, 49  temporal and geometric features are extracted from 5,603 signatures of 105 subjects. In newer contributions such as\cite{Fierrez-Aguilar:2005:OSV:2134859.2134923},  pressure is also used to compute features. Most approaches achieve an equal error rate (EER) between 1\% and 6\%~\cite{Jain20022963}.
The complexity of gestures differs between signature verification and touchalytics. Compared to  touch strokes, signatures are rather complex, which enables the extraction of more sophisticated features that support the authentication task.
The main differences are the tools used and the availability of data. While in touchalytics the recording of all kinds of raw features is implicit for this interaction and is therefore always available, in signature verification only the spatial features are available without explicitly augmenting either pen or paper with extra sensors. This renders touchalytics more suitable for continuous authentication in practice.
}

The two papers that are probably most related to our contribution are 
\cite{Sae-Bae:2012:BGN:2207676.2208543} and
 \cite{munichGuys}. Both methods try to match recorded touch data with historical touch data of the user. In \cite{munichGuys}, the authors augment  a gesture-based authentication method with a behavioral classifier that supports the authentication. Users that know the secret gesture cannot authenticate unless they carry it out in the very same way as the true user does it. 
In \cite{Sae-Bae:2012:BGN:2207676.2208543} a set of 22 multi-touch gestures are used to authenticate 34 users on an iPad. The authors achieve EERs of 7\%-15\% if the users performed one gesture, 2.6\%-3.9\% if two gestures were combined, and 3\% with one unique gesture for each user.
There are significant differences in the problem setting of these two papers and our contribution. While in \cite{munichGuys,Sae-Bae:2012:BGN:2207676.2208543}, a defined entry-point is required for the user to authenticate, we aim at an implicit and continuous scenario. Second, in our authentication scheme, the users can interact with the screen as they like, while in \cite{munichGuys,Sae-Bae:2012:BGN:2207676.2208543}  touch trajectories are  compared with a particular (secret) gesture.

\section{General Idea and Goals}

In this section, we provide a bird's-eye view on the idea of continuous authentication and set the goals of our study. In particular, we want to understand the scenarios in which such a mechanism would work reliably. 

\subsection{Continuous Touch-based Authentication}
The main hypothesis of this study is that continuously recorded touch data from
a touchscreen is distinctive enough to serve as a behavioral biometric.
Figure~\ref{fig:example} illustrates some strokes performed by different users
while reading text. These plots depict the $x$ and $y$ coordinates of each
stroke. In addition to coordinates, a commodity smart phone records times,
finger pressures, and the screen areas covered by each finger. A continuous
authentication application could run in the background and extract multiple
features from all available raw input. This raw input is readily available through the phone's API.  Based on various extracted features, the
system can then learn a profile of the legitimate user and compare all screen
interaction with this profile. There are two phases for learning and
classifying touch behavior.

\paragraph{Enrollment Phase}
Initially, the system must be trained in an enrollment phase. During that phase the system relies on a conventional authentication method, such as a 
 password challenge.

We define two particular user actions and call them `trigger-actions'. These actions should be frequent for any usage and primitive, i.e. they should be part of all more complex navigational gestures. Whenever the user performs a trigger action, the system logs the fingertip data. In our study, these actions involve:
\begin{itemize}
\item sliding horizontally over the screen. Usually, one does this to browse through images or to navigate to the next page of icons in the main screen.
\item sliding vertically over the screen to move screen content up or down. This is typically done for reading email, documents or web-pages, or for browsing menus.
\end{itemize}
We distinguish vertical strokes from horizontal strokes because 
it will prove easier to compare strokes within each trigger-action than across trigger-actions.
In principle, the set of trigger actions could be extended to more complex gestures, including multi-touch gestures like zooming. However, we focus on single-touch gestures as 
more  complex gestures are used too infrequently to be appropriate for continuous monitoring. 
As clicking exhibits  too few features to be discriminative for users, an authentication method must rely on the sliding actions.

During the enrollment phase, the system monitors the touch biometrics and
extracts particular features from the touch data (we will propose such features
in Section~\ref{touchalytics}). This process continues until the distribution
of touch-features converges to an equilibrium. This is the point in time when
one can assume that i) the user got used to her device and her device-specific
`touch-skills' no longer improve and ii) the system has observed
sufficiently many strokes to have a stable estimate of the true underlying
feature distribution of that user. At that point, the system can train the
classifiers and switch to the classification mode for authentication.

\paragraph{Continuous Authentication Phase}
Once the classifiers are trained,  the device begins the authentication phase. 
During this phase, the system continuously tracks all strokes 
and the classifier estimates if they were made by the legitimate user.
 For $t$ consecutive negative classification results, the system resorts back to the initial entry-point based authentication method and challenges the user.
Thereby, the precision of the individual classifiers influences the choice of
$t$. For a high precision classifier, only a few consecutive strokes suffice to
compute an estimate of the users authenticity; for a low precision classifier,
many strokes are needed. The choice of $t$ is proportional to the time
required to provide the first authentication decision. We will detail this in Section~\ref{sec_Nstrokes}.

The classifier strength affects the time it takes to make a decision. This
temporal dimension shapes the usage scenarios. For instance, if a few actions
suffice to provide a reliable classification, then an intruder can likely be
identified earlier, and potentially cause less damage. In this case one could
get rid of conventional password authentication, except for modifying the
security configurations of the device.  If the phone needs to monitor, e.g., an
hour of usage before giving a classification, our proposed mechanism could just
support the standard authentication mechanisms and serve as a theft detection
mechanism that responds to theft by activating {GPS}, sending {SMS},
or locking the device.

\subsection{\hi{Study Goals}}
We take the following approach to test the hypothesis that an authentication scheme that operates as described above is feasible.
We implement a proof-of-concept classification framework and challenge this framework with touch data that has been recorded from users interacting with different applications on a smart phone.
 We try to make the experimental conditions for collecting data as realistic as possible.

The main goal of our study is to analyze how robustly our proposed framework can distinguish users from each other. In particular, the  questions are: 
\begin{itemize}
\item What is the probability of rejecting a legitimate user?
\item What is the probability of accepting an attacker?
\item How long does the classifier need to make an authentication decision?
\item How robust is the classification within one session, across multiple sessions, and \hi{after one week}?
\end{itemize}

Our investigation aims at clarifying these questions and design decisions based on the technical feasibility of touch-based continuous authentication.

\section{Data Acquisition}
We carried out an experiment on Android phones where users must read text and compare images.
The main purpose of this experiment was to motivate users to produce many navigational strokes in a natural way.

\subsection{Experimental Protocol}
According to the protocol, the subjects were asked whether they wanted to
participate in an experiment on \emph{``reading and image viewing behavior on
smart phones''}. They were told to read three documents and to answer
comprehension questions after reading each document. They were also informed
that the given questions would be relatively easy.  The subjects were given a
mobile phone (We kept the number of
different mobile phones minimal to mitigate the influence of features based on
different technologies. As discussed in Section \ref{ExpDesign}, using many different phones may artificially improve accuracy and could invalidate the results). Next, a unique anonymous ID was assigned, a random
document selected and the participants were asked to start reading. After finishing
the document, a questionnaire was handed out with three multiple-choice
questions. This procedure was iterated for the remaining two documents.  In the
second phase of the experiment, the users were asked to spot differences in
pairs of similar images. 
 We stopped the
users after approximately two minutes, and repeated this procedure for the
second image pair. Finally, a questionnaire was handed out for general
statistics, and the participants were asked if they wanted to be part of a
follow-up study.

One week later, a follow-up study was conducted with only one document and one image comparison. Also, no general survey was handed out. Instead, after finishing all experiments, the participants were explained the true nature of the study.

\subsection{Recording Tool}
Collecting touch data from Android phones is limited by the fact that Android prohibits access of touch data across different applications, i.e.,  each application can only read touch data produced by interacting with the  application itself.
To conduct the protocol described in the previous section, enabling users to interact with the phone on various tasks, 
 we wrote an 
application for reading documents and viewing different images. It allows the entry of a
user ID, and contains links to the respective documents and image panels. For
the primary study, the application contained links to documents about
\emph{Wind}, \emph{Tulip Mania} and \emph{Yosemite National Park}. All of them
were excerpts from \emph{featured} Wikipedia articles. The images used to spot
differences are publicly available difference-comics (see Appendix).  
The user can see one image on one panel, and two screen-sizes away---separated by a black panel---is the second image. This means that users need at least two strokes to get from one image to the other. They were free to go back and forth as often as they wanted. 
 In the follow-up study a another difference
comic was selected, and we selected a new excerpt from a featured Wikipedia
article about \emph{Vampires}.  
There were no restrictions on the
orientation of the devices and users could switch freely between portrait and
landscape viewing modes.

During the experiments, the phones recorded the users' touch data. Sampled with
a variable frequency,\footnote{The devices log new touch events whenever a new
 pixel position is available. The sampling-$\Delta$ continuously ranged
from 1\,ms to more than 100\,ms, with a median of 17\,ms.} we recorded several
raw features:  an event code (e.g., \emph{finger up}, \emph{finger down},
\emph{finger move}, \emph{multi-touch}), the absolute event time in ms, and the
device orientation. For each present finger we recorded its $x$- and
$y$-coordinates, its pressure on the screen, the area of the screen covered by
the finger, the finger orientation with respect to screen and the screen orientation. All these are raw features that the Android system provides from a standard API.

For the primary study, the overall experiment time ranged between 25 to 50 minutes per subject. A single reading trial usually took between 10 to 15 minutes, while each image comparison trial took approximately 3 to 4 minutes. There were 41 participants in the study and four different smart phones with similar specifications were used (see  appendix).

\subsection{Experimental Design Decisions\label{ExpDesign}}
There are several possible pitfalls for proof-of-concept studies on authentication that can invalidate a positive outcome.  
Generally speaking, the challenge in carrying out such a study is to appropriately control all degrees of freedom such that the within-class variance is not artificially reduced and that the between-class variance is not artificially amplified. This means that the experimenter must ensure that the experimental setup enables a single user to operate on the device in any possible way. At the same time, it is important that all experimental conditions are the same for all users such that data from different users is free of differences that are caused by different conditions. Here, we address some of the issues we dealt with and explain how we tried to avoid them.

{\bf User adaption to experimental task.}
Users who know that the experiment serves to analyze their touch behaviors
might reflect about how to interact with the screen and, as a consequence,
behave in an altered way. Therefore, users were not told that we
analyze the way they used the touchscreen in our experiments. We revealed
the true purpose of the study only after the last touch data was collected.

{\bf Limited degrees of freedom.}
To avoid artificial limitation of intra-user variance, after reading a document we urged
the user to set aside the phone to fill out a sheet with multiple-choice
questions about the document. In this way, we provoke the user to randomly pick
up the phone, possibly causing the user to hold it in another way. As a result,
interaction with each document can be regarded as a new session.  Moreover, we
carried out follow-up experiments with different documents one week after the
first round.  We treat the touch logs recorded from one parts of the
documents as the enrollment phase and the logs recorded from other documents
serves to test the authentication method.

{\bf Influence of order of tasks.}
When every user reads all documents in the same order, then the order might have an effect on the touch behavior. Therefore, we generated uniformly random permutations and printed them on the protocol sheet  to provide the  reading order.

{\bf Adaptation over time.}
The touch biometrics of a user might change over time because users might improve the ability to control the device. In our study, we measure the touch behavior in a relatively short amount of time. This means we measure a snapshot of the users' behavior. 
In order to study long-term adaptation, one must give devices to each user and let them use it for a long time.  Due to time and money constraints, we refrained from carrying out such an analysis. However, we are aware that adaptation to the device might play a role and we asked each user how much experience with similar devices the user has.

In a real-world implementation of such an authentication system, there are two ways to deal with long-term adaptations of touch behavior. One way is to make the enrollment phase sufficiently long such that, presumably, the behavior converges to a stable state. The other way is to repeat enrollment phases from time to time.

{\bf Influence of phone and instructor.}
To prevent that different phone models and experimenters bias the results
towards a higher accuracy, ideally one must record all data with the same
experimenter and with the same phone.  However, we were curious about how the
phone or experimenter would influence the result, and to obtain a large number
of records, we distributed the workload between multiple experimenters.
Therefore, four different experimenters acquired the touch records and each
experimenter had their own Android phone. We used phones with similar screen
size and resolution (see Appendix). Moreover, we took care
that sufficiently many users were recorded on each phone to enable an analysis
of the influence introduced by the phones. See Section~\ref{sec_discussion} for
this analysis.

\section{Touch analytics}
\label{touchalytics}
In this section, we describe the features that we extract from the recorded data
and report on their statistics.

The first step of feature-extraction is to divide up the data records into individual strokes. A stroke is a sequence of touch data that begins with  touching the screen and ends with lifting the finger. One stroke $\mathbf s$ is a trajectory encoded as a sequence of vectors $\mathbf s_{n} =(x_n, y_n, t_n, p_n, A_n, o^{\text{f}}_n, o^{\text{ph}}_n ), n\in \{1,2,...,N \}$ with the location $x_n, y_n$, the time stamp $t_n$, the pressure on screen $p_n$, the area $A_n$ occluded by the finger, the orientation $o^{\text{f}}_n$ of the finger, and the orientation $o^{\text{ph}}_n$ of the phone (landscape or portrait).
Between two strokes $\mathbf s^{(m)}$ and $\mathbf s^{(m+1)}$, the touchscreen records no input.

We propose 30 features of a stroke all of which  are listed in Table~\ref{table:features}. Most of them are self-explanatory, but for some of them we like to detail how they are computed and why we believe they are relevant. The \textbf{coordinates of the two end-points} of the trajectory have been selected since we noticed that users tend to use distinctive screen areas   for their strokes. The  device reacts independently from the stroke location. Therefore, the choice of this location is completely left to the user and thus varies a lot over the users.

\begin{figure}[htb]
\centering
\includegraphics[width=0.5\columnwidth]{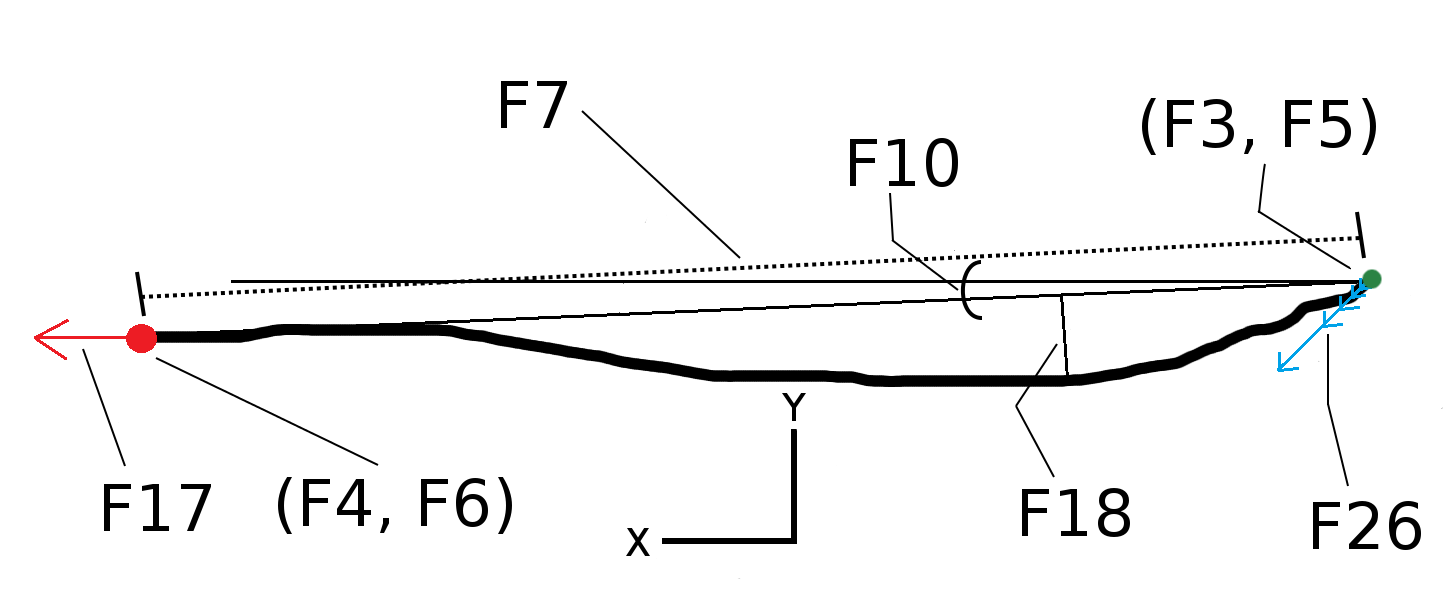} 
 \caption{Illustration of a few geometric features of a stroke (fat line). This  stroke starts off on the right side with the acceleration illustrated by blue arrows. It ends  at the red point with a finite velocity as indicated by the red arrow. The feature numbering equals the one used in Fig.~\ref{fig_corr}. 
 \label{fig_featureExample}}
\end{figure}

The \textbf{median velocity of the five last points} of the trajectory is able to distinguish users that stop the finger before lifting it from those that lift their finger while it still has a finite lateral velocity.  In the first case, the screen stays after the finger is lifted. In the latter case, the screen content gets `accelerated' by the finger and then continues moving even after the finger is lifted. Some users use this `ballistic' scrolling method, and if they do, they might have distinct throwing velocities.

The \textbf{mean resultant length} \cite{directionalStat} quantifies how directed the stroke is. All $N$ consecutive pairs $(x_n, y_n), (x_{n+1}, y_{n+1})$ define an ensemble of $N-1$ directions $z_n=exp(i\phi_n)$ with unit length and with angles $\phi_n$. 
The mean resultant length $R$ of this ensemble is characterized by $R=(N-1)^{-1} \vert \sum_{n=1}^{N-1} z_n \vert$. $R$ scales between 1 for a straight line and 0 for uniformly random angles of line segments. The  \textbf{mean direction} of the ensemble is $\arg ( (N-1)^{-1} \sum_{n=1}^{N-1} z_n )$.

We compute the \textbf{length of the trajectory} and the \textbf{direct distance between its end-points}. The ratio between these two defines another measure of angular dispersion. All strokes deviate a bit from the straight line. The \textbf{largest absolute perpendicular distance between the end-to-end connection and the trajectory} constitutes another feature. We project each vector on a perpendicular vector with a defined direction to distinguish if the largest deviation is on the left side or the right side of the end-to-end connection. This might be an indicator on whether the user is left-handed or right-handed.

An important factor in touch analytics is the time. Some users steadily and slowly scroll while reading. Others quickly scroll to a new position and  read on the still screen. This can be detected by \textbf{stroke duration} and \textbf{inter-stroke time}. Both are also informative about the reading speed which is supposedly different for different users.

We discard strokes with a too small displacement. Such strokes presumably represent single clicks.
All features that involve directional data are almost random for clicks and might confuse classifiers. 

\paragraph{Informativeness of Features}
In order to provide more insight in how user behavior differs with respect to individual features, we compute a measure of informativeness for each feature  $F$. We define this measure as the relative mutual information between the feature and the user~ID $U$ :
$$
I_F := \frac{I(F;U)}{ H(U)} 
= \frac{ H(U) - H(U \vert F) }{H(U) }
= 1 - \frac{H(U \vert F)}{H(U)}
$$
Here, $I(F;U)$ is the mutual information between the random variables encoding the feature magnitude $F$ and the user~ID $U$. $H(F)$ and $H(U)$ are the entropies of these variables.
For each feature, this measure takes a value between 0 and 1, whereas 0 means the feature carries no information about the user ID and 1 means the feature determines the user. To compute $I_F$, one must convert the features to discrete variables. We use 50 equally spaced bins that span a range from the 10\% quantiles to the 90\% quantiles of the features. This makes the range more robust against outliers than using bins that range from the smallest to the largest value.

\begin{figure*}[t]
\centering
\includegraphics[width=0.47\textwidth]{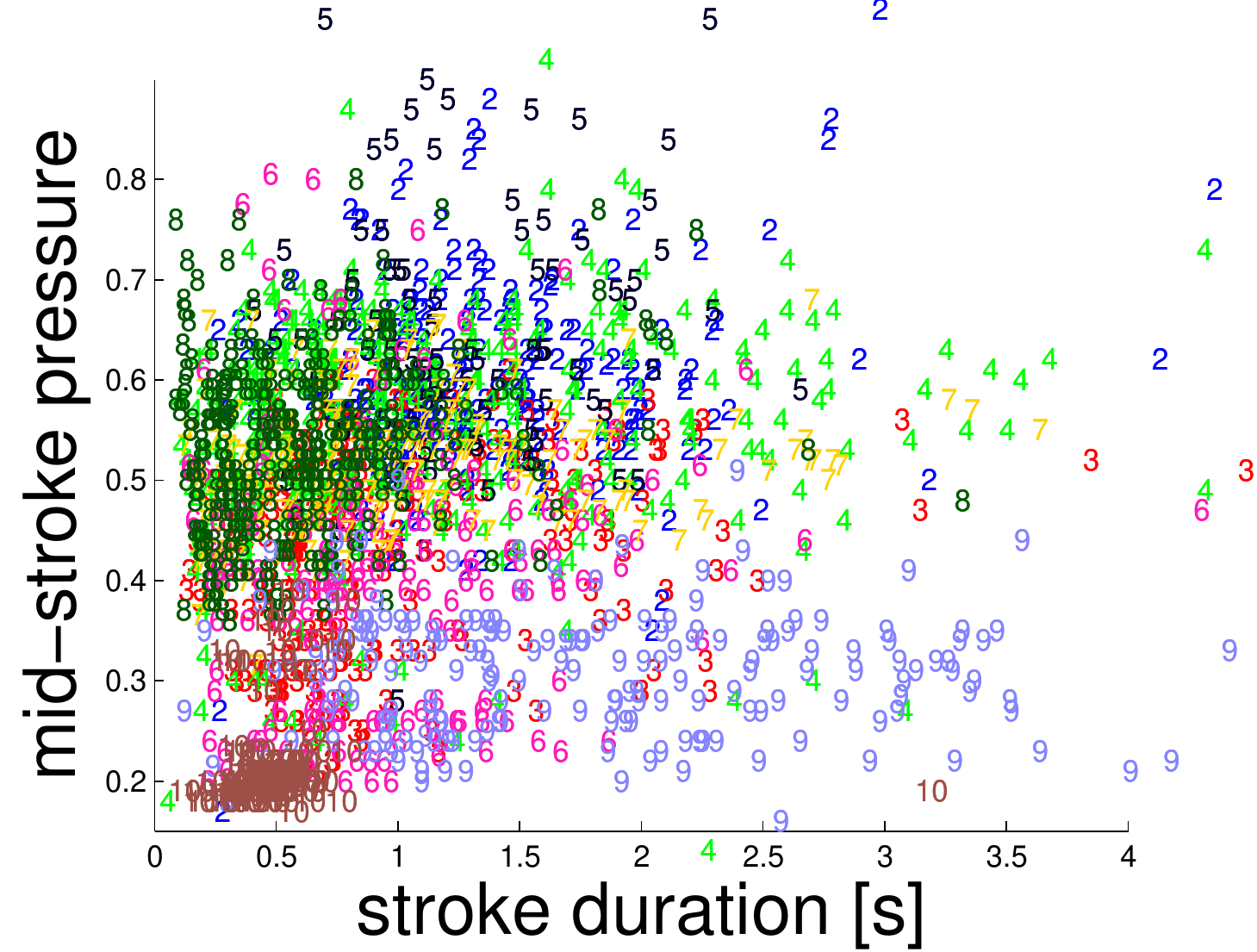}  
\ \ \ \ \
\includegraphics[width=0.47\textwidth]{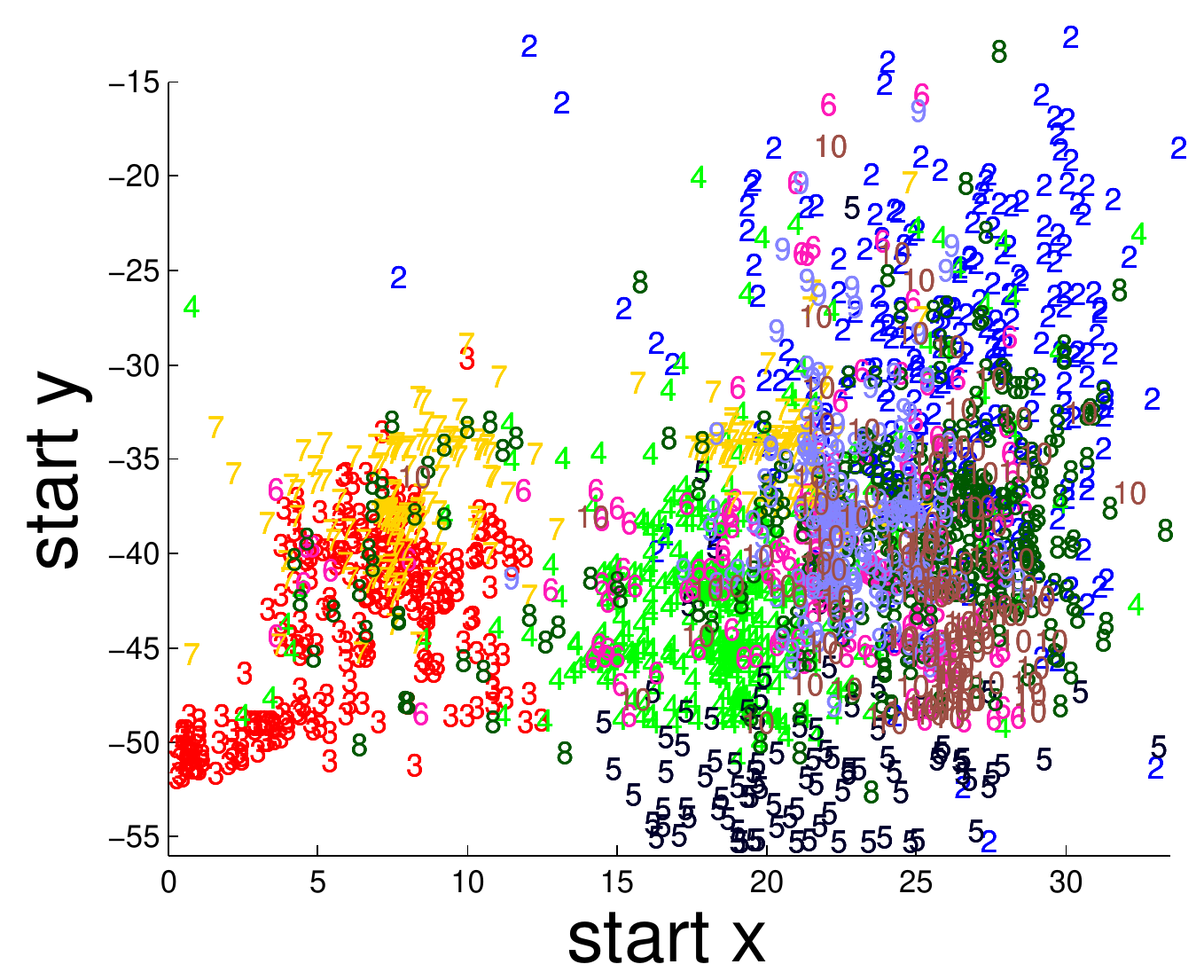}  
 \caption{Stroke features projected on a 2D-subspace. The user ID is given as a colored number. Already in these low-dimensional feature spaces, a class separation is apparent. The data depicted here was collected from users reading three Wikipedia articles in three different sessions. The left plot contrasts the finger pressure on the screen at the middle of the stroke against the stroke duration. The right plot shows the $xy$-positions where the fingertip first touches the screen.
\label{fig_2Dplots}}
\end{figure*}

\begin{table}[htb]
  \centering
\small
  \begin{tabular}{p{0.23\columnwidth} p{0.77\columnwidth}}
    \textbf{Rel. mutual information} & \textbf{Feature description}  \\ 
20.58\% & mid-stroke area covered\\
19.63\% & 20\%-perc. pairwise velocity\\
17.28\% & mid-stroke pressure\\
11.06\% & direction of end-to-end line\\
10.32\% & stop $x$\\
10.15\% & start $x$\\
9.45\% & average direction\\
9.43\% & start $y$\\
8.84\% & average velocity\\
8.61\% & stop $y$\\
8.5\% & stroke duration\\
8.27\% & direct end-to-end distance\\
8.16\% & length of trajectory\\
7.85\% & 80\%-perc. pairwise velocity\\
7.24\% & median velocity at last 3 pts\\
7.22\% & 50\%-perc. pairwise velocity\\
7.07\% & 20\%-perc. pairwise acc\\
6.29\% & ratio end-to-end dist and length of trajectory\\
6.08\% & largest deviation from end-to-end line\\
5.96\% & 80\%-perc. pairwise acc\\
5.82\% &  mean resultant lenght\\
5.42\% & median acceleration at first 5 points\\
5.39\% & 50\%-perc. dev. from end-to-end line\\
5.3\% & inter-stroke time\\
5.14\% & 80\%-perc. dev. from end-to-end line\\
5.04\% & 20\%-perc. dev. from end-to-end line\\
5.04\% & 50\%-perc. pairwise acc\\
3.44\% & phone orientation\\
3.08\% & mid-stroke finger orientation\\
0.97\% & up/down/left/right flag\\
0\% & change of finger orientation\\
  \end{tabular}
  \caption{List of extracted features. For each individual feature, we report the mutual information with the user ID.  Please see Figure~\ref{fig_corr} for the pairwise correlation coefficients. \label{table:features}}
\end{table}

The outcome of this analysis 
is depicted in Table~\ref{table:features}. The most informative single features are area covered by  fingertip, the 20\% percentile of the stroke velocity,  fingertip pressure on screen and the direction of the stroke. 
They are followed by the locations of the end-points of the trajectory. 
For scrolling, the $x$-positions of the endpoints are more informative than the $y$-positions. This reflects that users adjust the $y$-positions to the desired scrolling speed. The  $x$-position is completely up to their choice and thus only depends on the users' accustomed behavior.  The change of finger orientation provides no information gain. This might be due to the fact that the measurement of finger orientation is very insensitive and almost always takes the same number.

Please note that this ranking does not mean that the few topmost ranked features constitute the most informative collection of features. One can gain more information by combining features that complement each other. This also holds for some other strongly correlated pairs, such as the  80\%percentile of velocities and the 50\%percentile of velocities (the median), the length of the trajectory and the end-to-end distance, and the average direction of all segments of the trajectory and the direction of the end-to-end connection. 

\begin{figure}[htb]
\centering
\includegraphics[width=0.7\columnwidth]{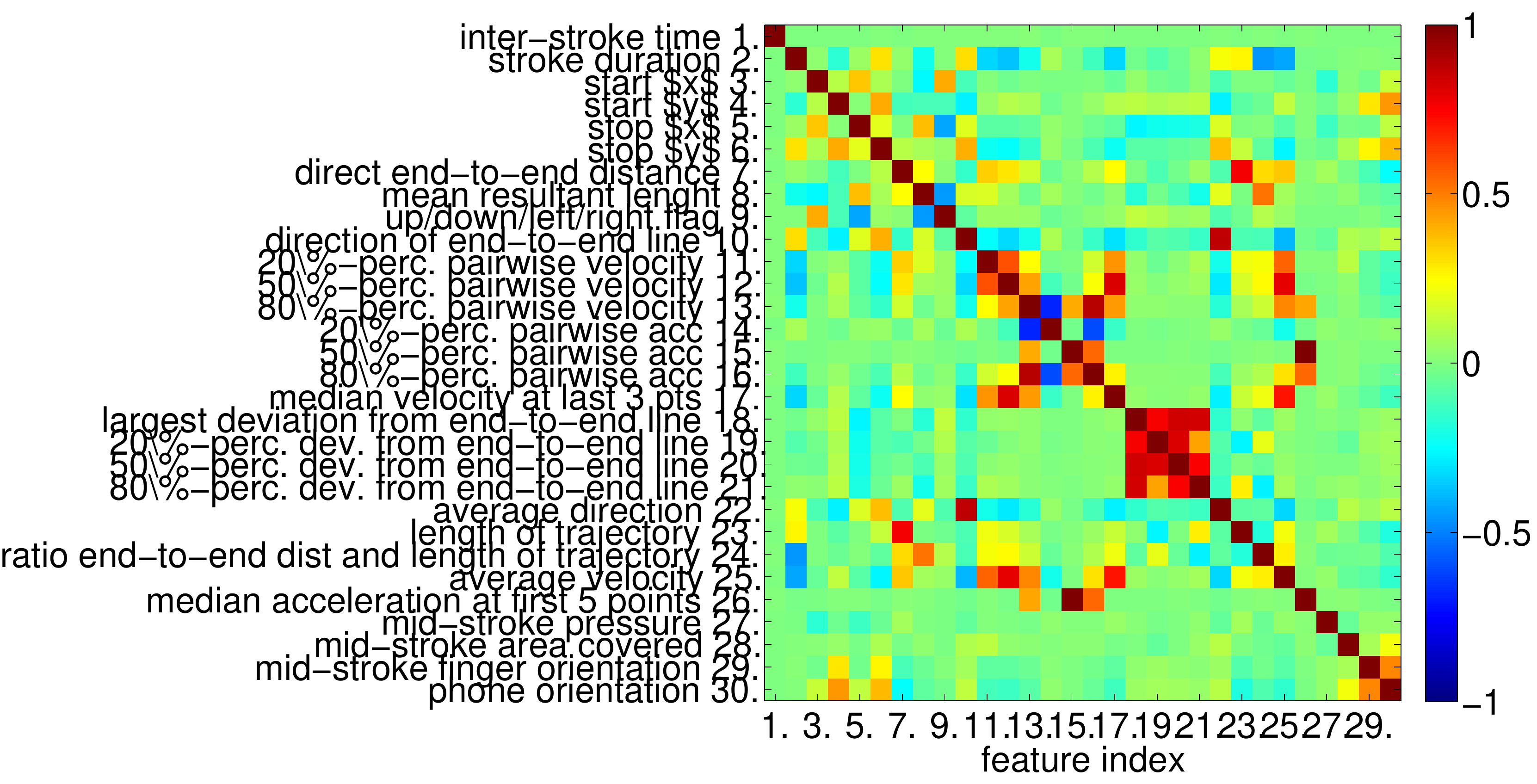}  
 \caption{Correlation matrix of all features. For some feature pairs with a strong correlation, we only use one of them  as detailed in the text.\label{fig_corr}}
\end{figure}

In order to better understand which features provide redundant information, we depict the correlation coefficients of all pairs of features in a color-coded plot in Figure~\ref{fig_corr}. Green encodes that the feature pair is not correlated, red indicates a positive correlation and blue indicates a negative correlation. The darker a color is, the larger is the absolute correlation coefficient. With this plot, one can spot a few highly correlated features. 
In general, it is risky to use correlation as a guideline for feature selection as two correlated features can still improve classification when they are in the same collection of features \cite{Guyon:2003:IVF:944919.944968}. Therefore, we use our understanding how the features are computed to discard individual features and thereby speed up the learning phases of the classifiers. The correlation coefficients merely serve as a confirmation and also to illustrate our decisions.
For instance, we identify that the length of the trajectory and the end-to-end distance are highly correlated. Semantically, they quantify almost the same thing. As   their ratio constitutes another feature anyway, we ignore the length of trajectory in the classification. Due to redundancy with the average direction, we also discard the orientation of the end-to-end line and the average velocity. 
 All the other features are used in all experiments without another feature selection step.

\section{Classification framework}
\label{sec_classifiers}
Empirically, the stroke features described in the previous section exhibit a larger variance across different users than for a single user. This motivates that a classifier can  distinguish different users according to their stroke features. In this section, we propose a framework for solving this task.

\paragraph{Choice of Classifiers}
We use two different classifiers, $k$-nearest-neighbors ($k$NN) and a support-vector machine with an rbf-kernel (SVM). Our decision for these classifiers was driven by various reasons. 
$k$NN  is robust to work with and provides a fast classification. The $k$NN classifier takes every new observation (here: a stroke) and locates it in feature space with respect to all training observations. The classifier identifies the $k$ training observations that are closest to the new observation. Then, it selects the label that the majority of the $k$ closest training observations have. This procedure requires no explicit training phase. The classifier merely stores all training observations and their labels. For huge datasets, the limitation of this method can be that not all data can be stored. In our case, this is not a problem as our feature space is comparably low-dimensional and, to keep classes balanced, we store only as many samples from the negative class as there are samples of the legit user.
The search time for the $k$ nearest-neighbors can be accelerated by computing a neighborhood relation of the training examples prior to testing.  We used a k-d tree to organize the training observations. In this way, the number of distance calculations for $n$ randomly distributed observations is reduced from $n$ to $\log(n)$ \cite{matlabimplemknn}. 
We used a Euclidian distance and selected the parameter $k$ from all odd numbers between 1 and 7 by cross-validation on the training data. Interestingly, $k$NN can naturally be used to solve a multi-class classification problem. This problem is harder than authentication. Only because we put all other users of the training set in the same (the negative) class, we turn the problem into a binary decision. 

Support vector machines \cite{svmCortes} are popular and powerful binary classifiers. 
In our authentication scenario the two classes are i) user of interest and ii) all other users. SVMs divide the feature space by a hyperplane such that the margin between the  two classes is maximized, i.e., SVMs squeeze a maximally thick hyper-brick between the boundary observations of both classes, the so-called support vectors. In contrast to $k$NN, SVM generalizes from the observed data, i.e., it forgets the individual observations after training and only saves the decision hyperplane.
 For more robustness against outliers, a small number of boundary observations are tolerated within the margin. A parameter $C$ controls the trade-off between maximizing the margin and minimizing the number of such exceptions.
For classes that are not linearly separable in feature space, one can replace the standard scalar products involved in the computation of the hyperplane with so-called kernels. Kernels implicitly transfer the problem in another high-dimensional space where the classes are separable. In the same step, the kernel maps the found hyperplane back to feature space \cite{Boser:1992:TAO:130385.130401}.
We use a Gaussian radial-basis function as the kernel, parameterized by the width parameter $\gamma$. We tune the two relevant parameters $\gamma$ and $C$ of this rbf-SVM by five-fold cross-validation on the training data. We expect that the SVM improves accuracy for borderline strokes whose feature space location is between user-classes.

\paragraph{Training, Testing, and Evaluation}
After feature extraction, we subdivide the dataset into a training set and a test set. Depending on the usage scenario, we use different ways for this subdivision. In Section~\ref{sec_results}, we explain this in more detail.

We normalize and standardize the data. The training set is used to train the classifiers and to tune their parameters. 
We then test the classifiers on data that has not been seen by the classifiers during training time. Our evaluation metric  involves the false-acceptance rate (FAR), the false rejection rate (FRR), and the median time $T$ required to make the first authentication decision in a session. FAR is the fraction of strokes of imposters that are recognized as strokes of the legit user by the classifier. 
 FRR is the fraction of strokes of legit users that are rejected by the classifier. FRR quantifies the empirical probability that the legit user must resort to conventional authentication mechanisms. Put in a temporal context, if $T_s$ is the average time between two strokes, then the expected time after which the legit user must type in a password due to misclassification is $\text{FRR}^{-1} T_s$. 

The two error rates FRR and FAR can be traded off against each other. At the cost of missing out some imposters one can reduce FRR by making the classifiers less sensitive. And at the cost of more false negatives one can increase the probability of detecting intruders. In order to account for this usability-security tradeoff, we report the equal error rate (EER) in all experiments. This is the error rate at the sensitivity of the classifier where FAR equals FRR.
In the training step, we use five-fold cross validation to tune all involved parameters such that the smallest EER on the training data is achieved. For SVM the parameters are $C$ and $\gamma$, for $k$NN the parameter is $k$.

\paragraph{Combining Multiple Strokes}
Our classifiers treat every stroke individually. The estimation of the authenticity of the user is thus a highly volatile random variable. However, this estimation can be rendered more robust by bundling several consecutive strokes and classifying them together. Instead of individually classifying all strokes and taking the majority vote as the final decision, we combine the classifier output at an earlier stage. For SVM, we average the continuous scores of projecting the individual test observations on the vector orthogonal to the decision hyperplane. The final classification is then the thresholded average score, depending on where to allocate the FRR against FAR trade-off. For $k$NN, we sum up the number of positive and negative labels of all nearest neighbors of all involved strokes and put the threshold on the ratio of these counts. In all experiments, we do not resolve the trade-off between FRR and FAR, as the preferences for one or the other clearly depends on the individual application scenario. To still account for this trade-off, we always report the equal error rate (EER).

\section{Experimental Results}
\label{sec_results}
We carried out various experiments to investigate the feasibility of continuous touch-based authentication.
In this section, we report the results of our experimental analysis.

\subsection{Influence of the Number of Strokes} \label{sec_Nstrokes}
\begin{figure}[htb]
\centering
\includegraphics[width=0.5\columnwidth]{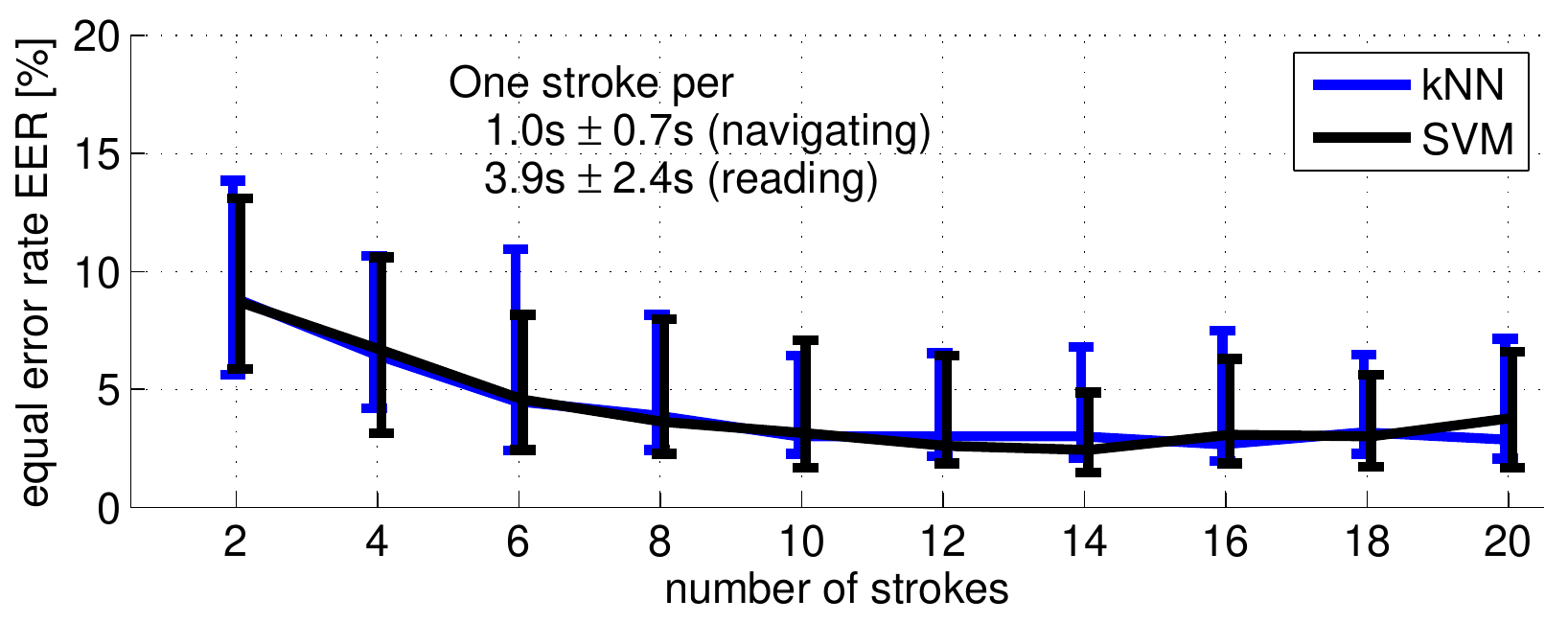}  
 \caption{The equal error rate (ERR) as a function of  the number of strokes taken for one classifier decision. 
 The error converges at 11 to 13 strokes. For the first decision, the system must wait until all required strokes have been recorded. For all subsequent decisions, a sliding window of past strokes will be taken such that a new decision can be made  at every other stroke. \label{fig_errvsNstrokes}}
\vspace{-2ex}
\end{figure}

The reliability of the classifier output depends on the number of strokes that are used to estimate the authenticity of a user. We analyzed the influence of this parameter by running several tests on inter-session authentication with the scrolling classifiers varying the number of strokes per decision. The outcome of this experiment is depicted in Figure~\ref{fig_errvsNstrokes}. 
When deciding with a single stroke only, the EER is approximately 13\%.
Both classifiers obtain a lower error when increasing the number of strokes used to provide a classification output. At a level of 11 to 12 strokes, the EER converges to a range between 2\% and 3\% and stays there up to using 20 strokes.

The choice of the number of strokes $n$ introduces another trade off between the robustness of the classification and the time needed to obtain the classification. While the robustness affects both the usability of the system (reduction of false rejections) and its security (reduction of false accepts), the time directly affects security, as it defines how long an attacker can interact with the device. \hi{For a given $n$, attacks carried out with the first $n$ or less strokes cannot be prevented}. Please note that $n$ only influences the time between turning on the device and the first decision of the classifier. For all consecutive decisions, a sliding window of the preceding  strokes is available for robust decision. Moreover, as the set of strokes used has a temporal order, one can adapt the number of strokes used based on the confidence of the classifiers. For instance, if the first three strokes show strong evidence for an impostor, one could make an early intervention, and if the odds are indifferent, one could wait for the next few strokes to come. In our experiments, the median user makes one stroke per 3.9 seconds while reading a text and one stroke per 1.0 second while navigating between images. In our experiments, we took 11 strokes for each decision, resulting in 11 to 43 seconds until the first decision is available.

\subsection{Application Scenarios}
\hi{Security risks and usability requirements of smart phones vary with context and environment~\cite{SPSM12_contextthreatmodels}. 
Touch-based authentication might provide enough security in some situations, such as using the phone at home, while being insufficiently secure for others (airport, restaurant, etc.). Also, it might be secure enough to increase the lockout time of standard PIN-protected screen-lock by a few minutes while it might be insufficient as the exclusive  security mechanism of a device. To reason about which implementations of the method provide a good trade-off between security and usability one must investigate its accuracy as a function of the time between enrollment and authentication.}
We experimentally analyze three different authentication scenarios. Each scenario corresponds to a different way of using continuous touch-based authentication and each one constitutes a different problem difficulty. With all data that we have collected, we can analyze the different scenarios by different ways of selecting training data and hold-out test data.
In the following, we describe all experimental settings.

{\bf \hi{Inter-week authentication}.}
The first scenario is the most challenging one. It assumes that the user trains the mechanism during an enrollment phase and then the classifier stays the same over many days up to weeks. When the user picks up the device, the mechanism must authenticate her with the classifier trained a week ago. We simulate this experiment by training the classifier with data that has been recorded over multiple sessions on the first round of data acquisition and test it on data recorded one week later.

{\bf Inter-session authentication.}
In the second scenario, the goal is to authenticate the user across multiple sessions at the same day. We train the classifiers on data that has been recorded on one or two sessions at the same day. Between every session, the user puts down the phone (to answer our survey questions) and picks it up again in a possibly different way. The test data is another session recorded at the same day. This is a scenario that would enable the user to use the device for a longer time without unlock between sessions. In this way, a password would still be needed to sign up every other day, but overall the usability would already be increased by making a frequent unlock step obsolete.

{\bf Short-term authentication.}
The last scenario is continuous authentication within one session of using the device. In this setting, the user authenticates by an entry-point authentication mechanism. Directly after authentication the device learns the stroke features of the current session. After having observed a few strokes, the device turns to classification mode and is possibly able to detect if another person takes the phone. This usage scenario cannot replace a password. However, one could imagine to use this as a complementing mechanism to extend secure authentication from seconds after typing the password to minutes. We simulate this scenario by randomly drawing training and test data from all available sessions.

\begin{figure*}[htb]
\centering
\includegraphics[width=0.45\textwidth]{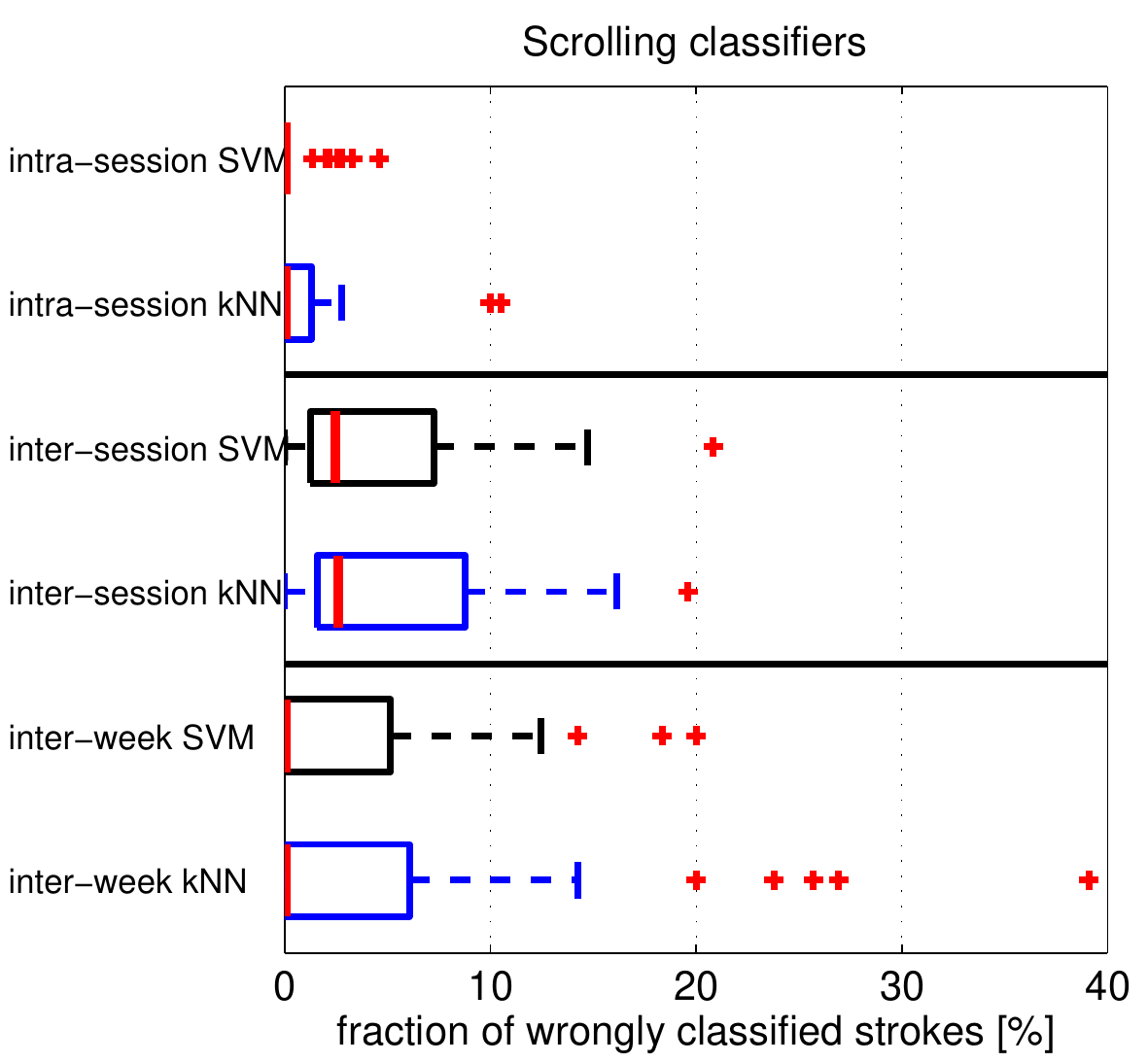}  
 \ \ \ \ \ \ \ \ \ \ \ \
\includegraphics[width=0.45\textwidth]{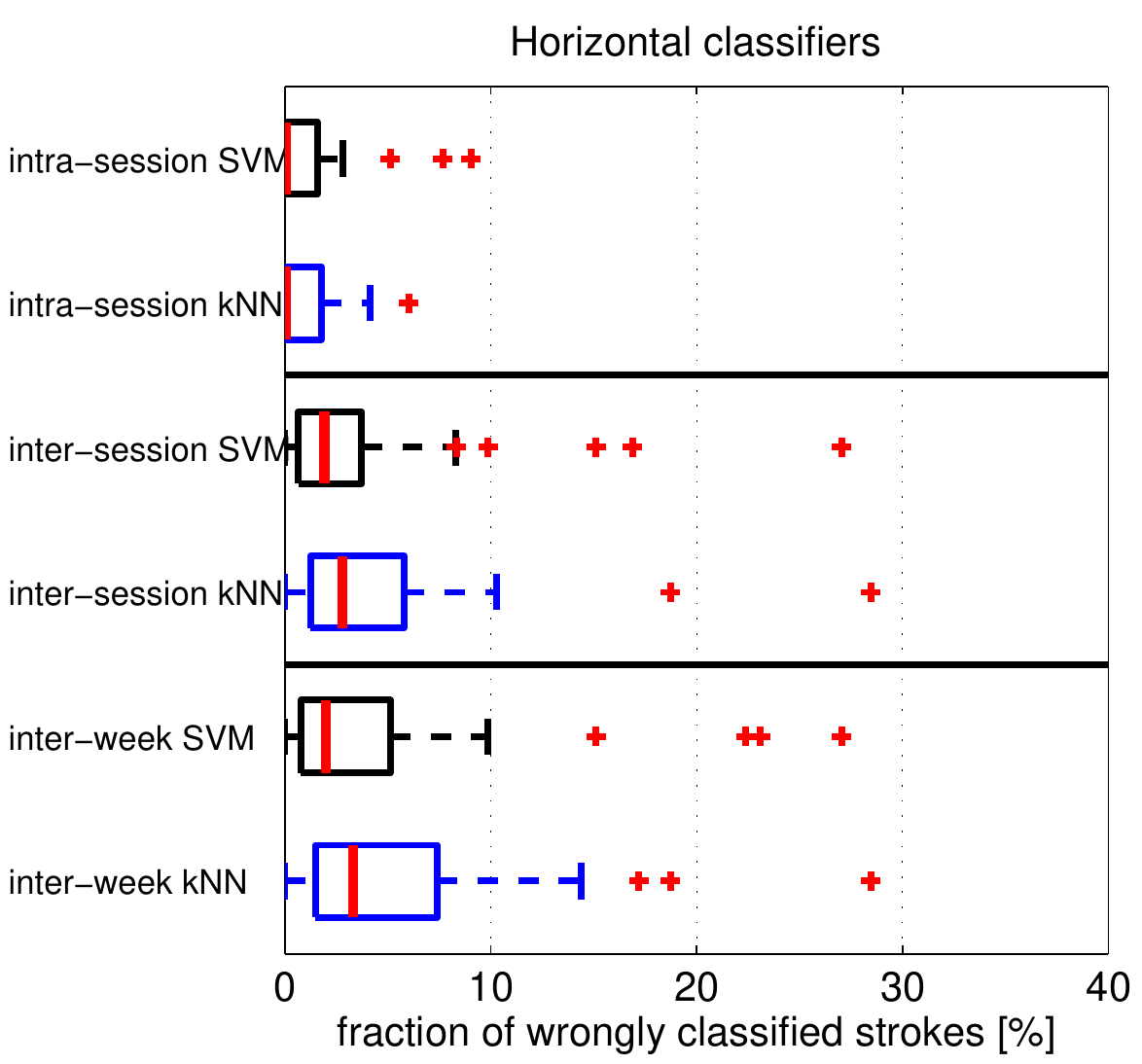}  
 \caption{ Summary statistics of equal error rates (EER) for scrolling classifiers (left) and horizontal classifiers (right).  The plots illustrate the performance in different authentication scenarios.
In `intra-session' experiments, the test set consists of randomly subsampled strokes from the entire corpus of strokes, such that the classifiers can learn from strokes of all sessions. In the `inter-session' experiments, the classifiers are trained on two sessions of the user and tested on another session.
 Users have put down the phone between sessions. In the `\hi{inter-week}' experiments, there is on week between recording the training data and the test data.  \label{fig_errholdoutv}}
\end{figure*}

{\bf Results.}
We depict the outcome of all three experiments in Figure~\ref{fig_errholdoutv}. The left plot illustrates the equal error rates (EER) for the scrolling classifiers and the right plot illustrates EER for the horizontal classifiers. All SVM results are colored in black, the results of $k$NN are colored blue. 
The boxplots depict the median EER (center red line) and the 25\% and 75\% percentiles, respectively. The whiskers span 1.5  times the inter-percentile distances. Outlying users with an error outside the whiskers are individually reported as red crosses.

The median EER ranges from 0\% to 4\% across all usage scenarios. The median intra-session errors are 0\%, whereas few outliers can reach a 10\% EER. It seems that, within one session, most users do not considerably change their touch behavior. The inter-session EER reaches from 2\% to 3\% and the inter-week EER reaches from 0\% to 4\%, depending on the scenario and the classifier used. The SVM achieves always a lower error than the $k$NN method. 
The EER quantifies the classifier configuration that results in equal false rejection rate (FRR) and false-acceptance rate (FAR).
Depending on the security requirements of the scenario, one could further reduce FRR the at the costs of a larger false-acceptance rate FAR.

Overall, the authentication difficulty seems to increase with increasing temporal distance to the training phase. Interestingly, the \hi{inter-week authentication} of the scrolling classifiers is an exception as its median error rate is lower than for the inter-session authentication. We interpret this result as an artifact introduced by the different document lengths in the first round of the experiment and in the second round. In the first round the subjects needed 25-50 minutes to read through all the texts. Most subjects found the text boring and some complained about the length of the experiments (one subject even aborted it). To convince users to participate in a second round one week later, we used a significantly shorter text. While in the first round, users had lots of time to hold the phone in many different ways, the short time of the second round experiment could have limited their interaction to their preferred way of holding the phone, with a resulting smaller intra-class variance of the recorded data simplifying the classification problem.
Yet, the large number of outliers for the inter-week experiment indicates a high difficulty for at least a few users.
For  gaming app data (horizontal classifiers), in every scenario users optimized the way to hold the device to best solve the image comparison task.

\section{Critical Discussion, Limitations and Extensions}
\label{sec_discussion}
Our experimental results suggest that it is possible to distinguish users 
 based on the way how they perform low-level interactions with a touchscreen. Thereby, depending on the authentication scenario, there is approximately a 0\% to 4\% chance that the correct user will be rejected or that a false user will be accepted. For some scenarios, this error rate is still too high for the system being directly implemented as is. However, this result demonstrates that touch-based continuous authentication is feasible. Our future research will aim at pushing the equal error rate lower. In this section, we discuss avenues for improving the accuracy of our proposed authentication scheme. At the same time, we highlight critical points and pitfalls in the design of our experiments and point to  limitations of our method.

{\bf How can the method be extended to other input?}
 One way to improve the results would be to take usage context into account. 
  For instance: the feature space could be extended by a categorical variable that takes values such as `read email', `navigate', `write email', `browse', `read pdf', `control music/video player'. This has two consequences: i) the classifier operates conditioned on the scenario (probably users behave differently in different scenarios) and ii) the scenario itself provides a soft evidence about the user identity. (for instance: if a user never uses video, it is suspicious if, at some point, video is heavily used).

\hi{
{\bf Application scenarios in light of temporal instability.}
Our inter-week experiments in Section~\ref{sec_results} suggest that the current method cannot securely serve as an exclusive authentication mechanism of a device. A satisfying usability-security trade-off might be achievable in short-term authentication scenarios. The touch data of a contiguous session after a PIN-login can be used to extend the screen-lock time for a few minutes. Also, a context-adaptive system could sacrifice false-negative performance in favor of a better usability when the user is in a private environment as estimated on GPS data and available wireless networks. 
}

{\bf Can the method also be applied to tablet computers?}
As we carried out our study on smart phones, an interesting question is if our findings do also apply to tablet computers. However, we refrain from such claims as this would be  speculation. In fact, we believe there are differences between smart phones and tablet computers that might make it harder to continuously authenticate users on tablets. In particular, we believe that the small size of the screen of smart phones helps continuous authentication. The reason is that content of documents, emails, image collections, menus, or icon collections hardly fit on the smart phone's screen in most application scenarios. As a result, the user must move around screen content  and thus the classifier gets a lot of observations over time. In contrast, on large tablet screens users can read for a long time without scrolling, all icons fit on screen, and so on. This might reduce the strokes per minute below a rate that can be considered secure. Moreover, the large screen introduces more degrees of freedom.

{\bf Influence of sample size.}
Generally, a limited number of observations affect the precision of empirical estimates of a random variable. In particular, the performance of multi-class classifiers is biased towards better accuracies if the classifier is trained and tested on a small number of classes. In principle, we have a binary classification task. However, the variability within the negative class (the `other' users) is clearly affected by the number of users.

\begin{figure}[htb]
\centering
\includegraphics[width=0.6\columnwidth]{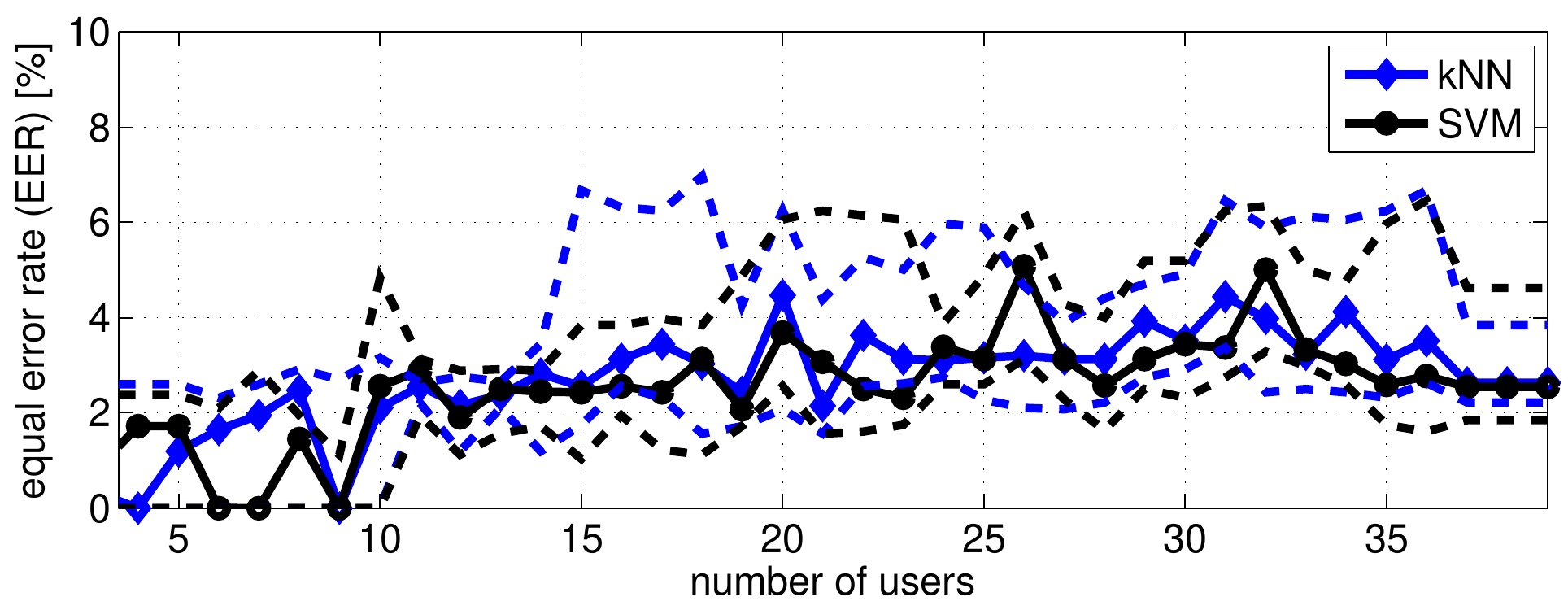}  
 \caption{ How accurate are our error estimates given that we test the classifiers on a limited number of users? This plot depicts the equal error rate of the inter-session experiment as a function of the number of subjects used for the experiment.\label{fig_NuserInfl}}
\end{figure}

 In order to investigate the influence of the number of subjects on the authentication error, we repeat one of the experiments, the inter-session authentication via scrolling classifiers, with a varying number of subjects. For each such number, we repeat the experiment ten times with a different random collection of subjects. For one such collection, we run three repetitions, each time with a different hold-out test session. The median equal error rates of this analysis, together with the 25\% percentiles and 75\% percentiles are depicted in Figure~\ref{fig_NuserInfl}. One can see that in the interval between 3 and 20 users the EER increases. But for more than 20 users, only small fluctuations within the error range are apparent. This demonstrates that our sample size is located in a range where it's influence is negligible. 

{\bf Influence of phone differences and instructor differences.}
We recorded data at three different sites, by four different experimenters, each with an own Android phone.
 For a discriminatory analysis, this bears the risk that the different experimental conditions alter the data such that it is easy for classifiers to distinguish all records from each other. As a consequence, a method that works well on such data might just be a good classifier for the type of device on which the data was recorded and not necessarily for the user. For instance, the screen of different phones has slightly different dimensions, affecting the number of dots per inch. We convert all data to relative values and normalize. But there might still be an influence. Another source of a device signature could be different levels of stickiness of the screen, different instructions given by the device owner, etc.

We are well-aware of this problem and address it by minimizing the differences in experimental conditions as much as possible. Before collecting the data, we agreed on an experimental protocol and tried to strictly follow this protocol.
As a sanity check we investigate the influence of the device on the classification error by comparing experiments carried out on multiple devices with experiments carried out on single devices. In total, there are three experiments: experiments on the same phone with the same instructors, experiment on the same phone make (Nexus One) but differing instructors, and experiments on all phones with all instructors. For a fair comparison, we constrained the number of users to the smallest number of users that was recorded with each device respectively.

\begin{figure}[htb]
\centering
\includegraphics[width=0.5\columnwidth]{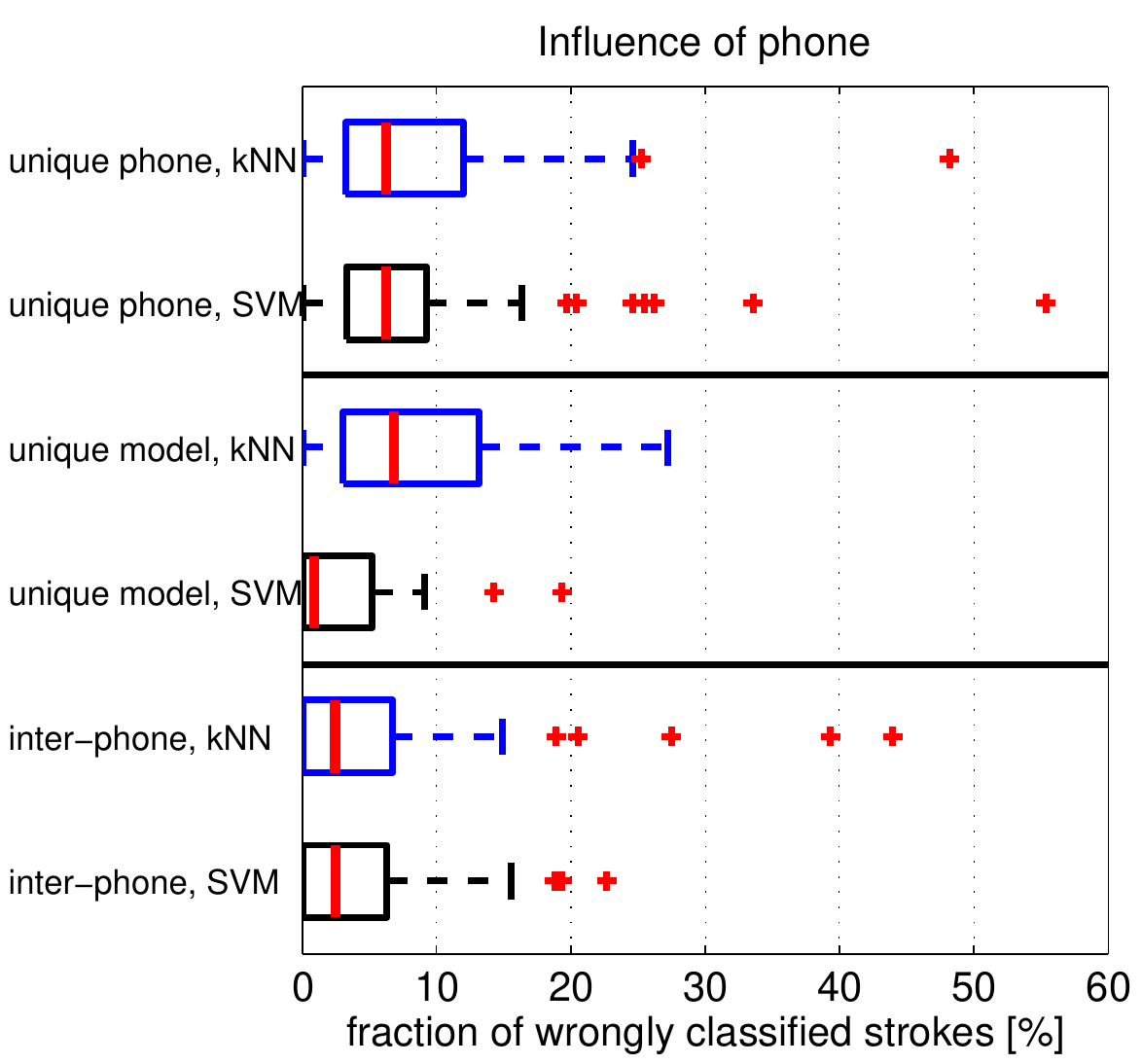}  
 \caption{Does the phone or the experimenter introduce a signature in the touch behavior? Then classification would be easier for strokes that have been recorded on different phones than for strokes recorded on the same phone.   \label{fig_PhoneInfl}}
\end{figure}

The results of this sanity check are illustrated in Figure~\ref{fig_PhoneInfl}. In fact, the error rates for users on the same phone are on average 2\% higher than for user data collected on multiple phones. Thereby, it is unclear if the instructor introduces a signature that helps to distinguish users or if the device itself is responsible. This can be seen by setting the results in relation to the outcome of the experiments where the phone was the same make but the instructors differ (unique model). For these experiments, the $k$NN classifiers perform equally well as on data collected by a unique instructor suggesting that  the phone is responsible for the difference in error rate. However, the SVM has an error rate as low as for the inter-phone experiment, suggesting that it is the instructor who introduces a distinct signature in the touch data. This inconsistency points at an alternative explanation. Given the low number of users in this sanity-check experiment (on one phone we collected at most 16 users), the found differences could as well be due to the small sample size. We conclude that neither the influence of the instructor nor the influence of the phone alone is as large that it can be reliably detected with the given number of samples. The combined influence could play a role within the precision at which we can estimate the error rates. Therefore, conservative estimates should not interpret our experimental findings below a 2\% resolution.

{\bf Targeted attack versus random attack}
The false acceptance rates that we compute are based on users that do not actively try to mimic the touch behavior of another user. As a consequence, these error rates correspond to a random attack or to the attack of an uninformed attacker. More sophisticated attackers might try to observe the user's behavior to mimic it. However, except for the $xy$-coordinates of the stroke, we can hardly imagine someone learning the touch behavior of 30 features, such as pressure on screen, distribution of acceleration, etc., just by looking over the shoulder. A more successful but also more involved attack would be to place a malware application on the user's device. This malware could learn and report the touch pattern if the details of how to compute the features are known to the attacker. Such an attack might in fact have a very high success chance. However, we argue that a user with malware on the device has already lost the race against the attacker. To condition an attack on such a situation, renders the attack successful by construction.

\section{Conclusion}
We investigated the question of whether and how touchscreen input could serve as a behavioral biometric for continuous authentication. Importantly, we justified that simple touch movements, which are usually a part of any navigation activity, are sufficient to authenticate a user. Along this way, we carefully designed a data acquisition experiment to collect natural touch behavior of 41 subjects. 
We designed a proof-of-concept classification framework that extracts 30 different behavioral features from the raw touchscreen interaction data. The framework trains user profiles based on vertical and horizontal strokes using a $k$-nearest neighbor classifier and a Gaussian rbf kernel support vector machine. These classifiers achieve robust authentication results, with equal error rates between 0\% and 4\%, depending on the application scenario. The results suggest that our proposed method is applicable in a variety of scenarios that benefit from continuous authentication based on natural navigation gestures.

In summary, the aim of this work was to provide the first grounds on touch analytics. We see several avenues for future work. One way to further improve accuracy could be
the use of multi-stroke based features. As detailed in our extended discussion section, we will analyze how the dimensions of tablet computers affect touch analytics. Moreover, it is interesting to identify and resolve all design decisions for embedding our method in an actual system, possibly using multiple modalities.
Combining touch analytics with other modalities such as, for instance, location, accelerometer data, images from the front-facing camera, and application usage patterns promises an improved accuracy.

\section*{Acknowledgment}
This research was supported by Intel through the ISTC for Secure
Computing and by the Swiss National Science Foundation (SNSF), grant no.~138117.

\bibliographystyle{plain}

\medskip
\begin{appendix}

\section{Appendix}

In this section we provide background information on data collection and experiments that serves to further illustrate details of our experiments.

\subsection{Smart phones used}
\hi{We used five different smart phones of four different kinds, each with an Android system and with similar screen sizes. All phones operated on Android 2.3.x. .
Two experimenters (those with Exp-ID A and B) had a Nexus One phone. One of them also used a Droid Incredible phone to collect data from four users. All phones are listed in the table below.
}

\begin{table}[htb]
\centering
\begin{tabular}{l l l l l l}
Name & Screen & Resolution & Ratio & $\hi{N_{\text{users}}}$ & \hi{Exp-ID}\\
\hline
Droid Incr.& 94mm & 480$\times$800 px  & 252 ppi & \hi{4} & \hi{A} \\
Nexus One & 94mm & 480$\times$800 px  & 252 ppi & \hi{6} & \hi{A}\\
Nexus One & 94mm & 480$\times$800 px  & 252 ppi & \hi{13} & \hi{B}\\
Nexus S & 100mm & 480$\times$800 px  & 233 ppi& \hi{15} & \hi{C}\\
Galaxy S & 100mm & 480$\times$800 px  & 233 ppi& \hi{3} & \hi{D}
\end{tabular}
\label{tab:phonespecs}
\end{table}

\subsection{User Statistics}
 We collected data from 41 users recruited from students, members of our research groups, and neighbors. Most users (84\%) were right-handed, and 16\% were left-handed. However, during the experiments we noticed that users sometimes change their hands even within one session. Out of all users  32\% were female and 68\% were male.The table below provides the age distribution with a resolution of 10 years.

\begin{table}[htb]
\centering
\begin{tabular}{l l l l l l l}
$<10$ & 10-19 & 20-29 & 30-39 & 40-49 & 50-59 & 60-69 \\
\hline
0\% & 3\% & 78\% & 11\% & 3\% & 0\% & 5\%
\end{tabular}
\label{tab:userage}
\end{table}

\subsection{Image Comparison Game}
\begin{figure}[htb]
\centering
\includegraphics[width=0.39\columnwidth]{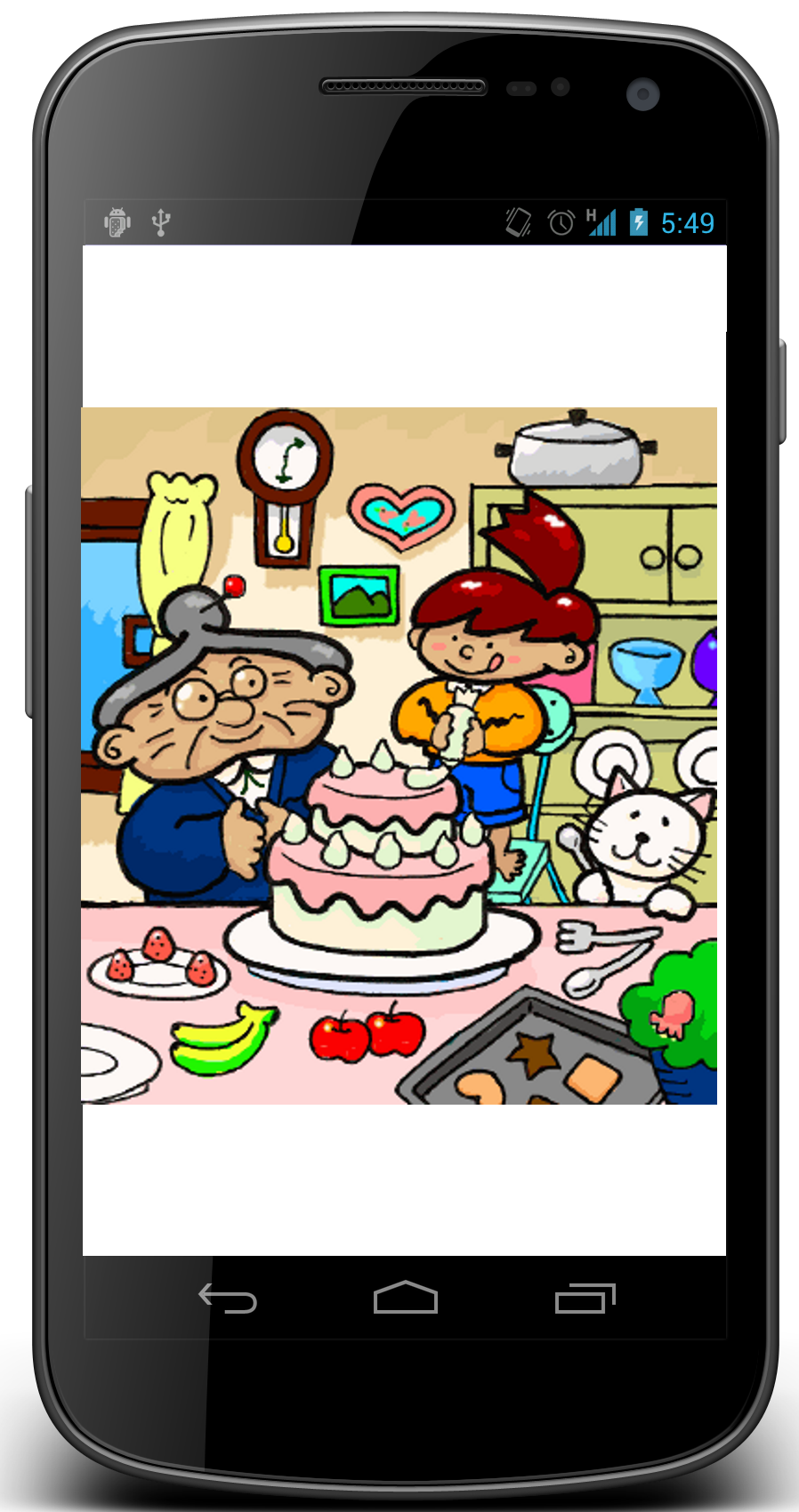} 
\hspace{36pt}
\includegraphics[width=0.39\columnwidth]{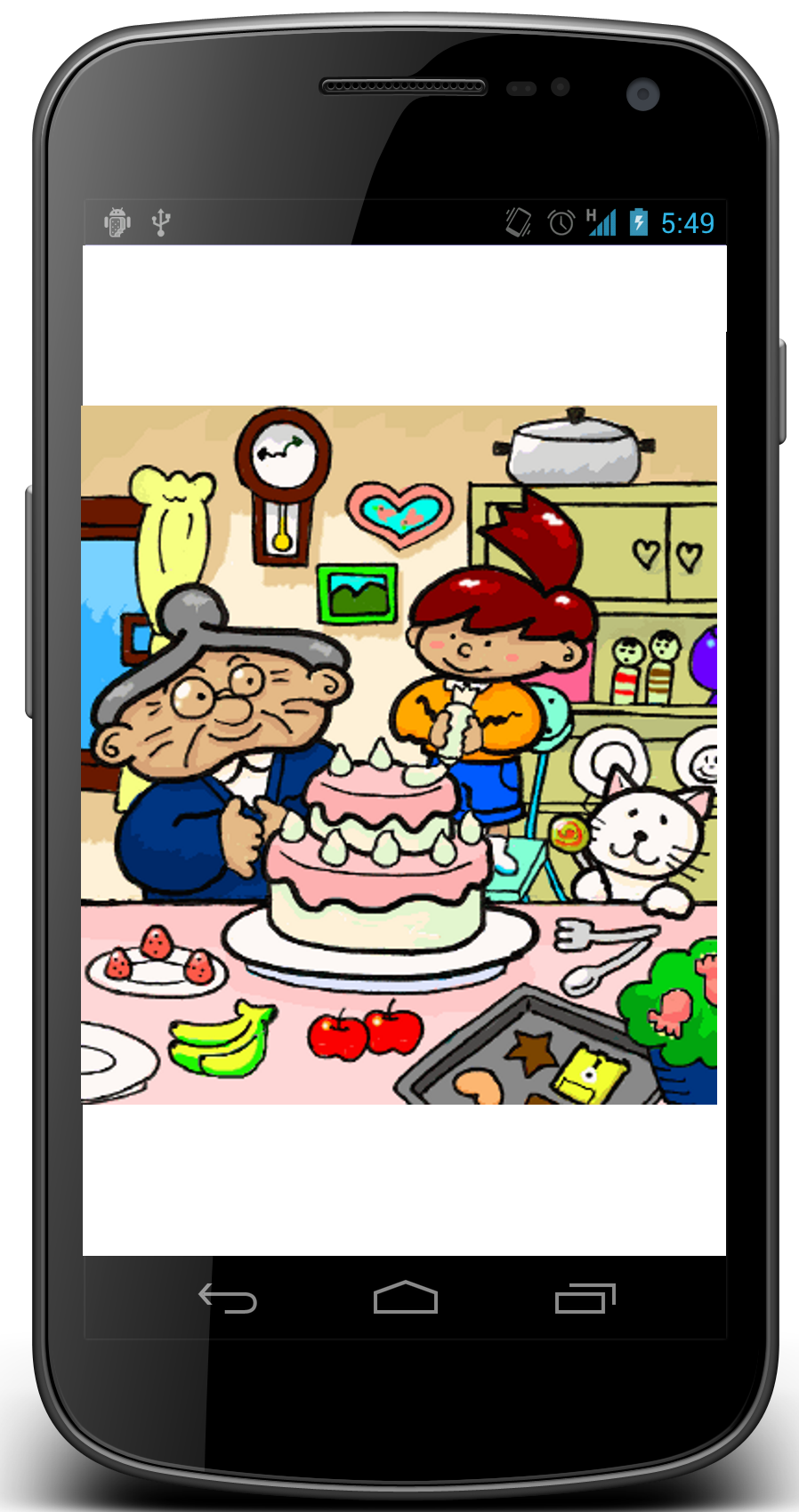} 
 \caption{Image comparison game. The two images exhibit many subtle differences. In order to go from one image to the other, the user must swipe away the screen content to the left or right. In order to prevent  the user from  seeing both images at the same time, they are separated by a black screen. \label{fig_imgExample}}
\end{figure}

Figure~\ref{fig_imgExample} depicts one pair of images that has been used for the image comparison application. Users had to move the screen content away to navigate from one image to the other. We designed the application such that it is impossible to see parts of both images at the same time. Most users found 8 differences on this example. The high score is 11 differences.

\end{appendix}

\end{document}